\documentclass[aps,pre,twocolumn,groupedaddress,showpacs]{revtex4}
\usepackage{graphicx}
\usepackage{amssymb}

\begin{document}


\title{Penetration depth for shallow impact cratering}


\author{M.A. Ambroso$^{1}$, C.R. Santore$^{1}$, A.R. Abate$^{1,2}$ and
D.J. Durian$^{1,2}$}
    \affiliation{
    $^{1}$Department of Physics and Astronomy, University of
    California, Los Angeles, CA 90095-1547, USA\break
    $^{2}$Department of Physics and Astronomy, University of
    Pennsylvania, Philadelphia, PA 19104-6396, USA
}


\date{\today}

\begin{abstract}
    We present data for the penetration of a variety of spheres,
    dropped from rest, into a loose non-cohesive granular medium.
    We improve upon earlier work [Uehara {\it et al.} Phys.\
    Rev.\ Lett.\  {\bf 90}, 194301 (2003)] in three regards.  First,
    we explore the behavior vs sphere diameter and density more
    systematically, by holding one of these parameters constant while
    varying the other.  Second, we prepare the granular medium more
    reproducibly and, third, we measure the penetration depth more
    accurately.  The new data support the previous conclusion that the
    penetration depth is proportional to the $1/2$ power of sphere
    density, the $2/3$ power of sphere diameter, and the $1/3$ power
    of total drop distance.
\end{abstract}



\maketitle




The mechanics of granular media continue to defy our intuition.  In
spite of their ubiquity in everyday life and industry, we have no
fully reliable rules for predicting response to an applied
force~\cite{brown,jnb,duran}.  If the forcing is weak, the medium remains at
rest and the local disorder in the packing gives rise to ramified
force chains with structures much larger than the grain size.  If the
forcing is strong, the medium can flow.  But when will the medium
yield?  Will the flow be smooth or intermittent?  How do velocity and
density vary with position and time?  No experimental characterization
currently exists that can be used to predict response in all other
sample and forcing geometries.  For example, flow down an incline
offers little insight as to how the same medium would flow between
rotating cylinders or on a vibrated plate.

Recently we investigated the mechanics of impact by projectiles
dropped into granular media~\cite{jun,katie}.  This is a situation
of natural interest.  Some of us would like to understand how far
our feet sink into the sand when walking, running, or jumping at
the beach.  Others of us would like to understand the lie of our
golf ball in a sand trap.  Still others would like to know how far
a warhead can penetrate the earth prior to detonation~\cite{rob}.
It's also interesting to consider the effect of impact on the
medium itself: the nature of the granular
splash~\cite{siggi,detlef2} and the morphology of the resulting
crater~\cite{roddy,melosh,holsapple,amato,hans,deBruyn,zheng}. Our
motivation is more general: to find a non-contrived situation
permitting the unusual nature of granular mechanics to be both
highlighted and characterized.  Projectile impact is ideal on both
counts.  It's interesting that penetration is nonzero even for
near-zero impact speed, but grows only very slowly with projectile
energy.  The average stopping force $\langle F \rangle$ can be
very small, but can also increase dramatically for stronger
impacts. This unusual mechanics can be studied quantitatively from
the penetration depth $d$ via a simple statement of energy
conservation:
\begin{equation}
    \langle F \rangle = m g H / d,
\label{econ}
\end{equation}
where $m$ is the projectile mass, $g=9.8$~m/s$^{2}$, and $H$ is the
total drop distance.  Note that $H$ is the sum of the free-fall height
$h$ and the penetration depth $d$ (see inset of Fig.~\ref{Raw}).

In Ref.~\cite{jun} we measured the penetration of spherical
projectiles of various densities $\rho_{b}$, and diameters
$D_{b}$, into loose non-cohesive granular media of various
densities $\rho_{g}$ and angles of repose $\tan^{-1}\mu$.  In all
cases the minimum free fall height was nearly zero, and the
maximum penetration depth was comparable to the ball diameter. All
of our data collapsed according to an empirical scaling relation,
\begin{equation}
    d = 0.14{1\over\mu}\left( \rho_{b}\over\rho_{g} \right)^{1/2}
    {D_{b}}^{2/3} H^{1/3}.
\label{junlaw}
\end{equation}
In Ref.~\cite{katie} we showed how this naturally generalizes to
cylindrical projectiles, independent of the tip shape.  If true,
Eq.~(\ref{junlaw}) has several interesting implications.  First,
it implies via Eq.~(\ref{econ}) that the average stopping force is
proportional to the tangent of the repose angle, $\mu$, consistent
with the notion that it represents a friction
coefficient~\cite{brown}.  Second, it implies that the granular
medium can be extremely fragile, suffering a nonzero penetration
even for zero free-fall height.  Eq.~(\ref{junlaw}) gives this
minimum penetration as
\begin{equation}
    d_\circ=(0.14/\mu)^{3/2}(\rho_b/\rho_g)^{3/4}D_b.
\label{juno}
\end{equation}
The penetration depth formula then can be recast dimensionlessly
as $d/d_\circ=(H/d_\circ)^{1/3}$.  Third, since Eq.~(\ref{junlaw})
is dimensionally complete, it suggests that the effects of
grain-grain cohesion and interstitial air are both negligible.  If
they were not, then even further physics would have to enter to
cancel the extra units. Air and cohesion effects can also be ruled
out because we found identical penetrations for granular media
that are identical except for particle sizes \cite{jun}. This is
to be expected, according to the Geldart classification scheme of
fluidization behavior vs particle size and density
\cite{gidaspow}. Fourth, and perhaps most curious,
Eq.~(\ref{junlaw}) implies that the penetration is not a function
of either impact energy, $\sim\rho_{b}h$, or impact momentum,
$\sim\rho_{b}h^{1/2}$.

The penetration of projectiles into granular media has also been
measured recently by other groups
\cite{ciamarra,deBruyn2,detlef3}. Ciamarra {\it et al.}
\cite{ciamarra} performed quasi-two dimensional experiments in
which a steel cylinder was dropped sideways into a packing of
rods.  The impact speeds varied by about a factor of five, and the
penetration depths varied from about 1.5 to 7 times the projectile
diameter. They report that the projectile deceleration is
time-independent and proportional to the impact speed. This
implies that the stopping time is constant and that the
penetration depth is proportional to the impact speed. de~Bruyn
and Walsh \cite{deBruyn2} performed experiments in which two
different diameter steel spheres were dropped into glass spheres
of five different bead sizes.  The impact speeds varied by about a
factor of four, and the penetration depths varied from about 1.2
to 5 times the projectile diameter. They report that the
penetration depth is linear in impact speed, but with an intercept
$d_\circ$ that can be positive or negative. This is modelled in
terms of a Bingham fluid, where the granular medium exerts a force
on the projectile according to a yield stress and an effective
viscosity, $F(v)=-F_\circ-bv$. Negative intercepts for depth vs
speed are predicted by this model. Even more recently Lohse {\it
et al.} \cite{detlef3} performed experiments in which spheres are
dropped at zero free-fall height ($h=0$), just barely touching the
sand. The projectile densities varied widely, at fixed diameter,
giving penetration depths $d_\circ$ ranging from about $1/4$ to 6
times the ball diameter. They report that the minimum penetration
is linear in projectile density. This is modelled in terms of
Coulomb friction, where the medium exerts a force on the
projectile proportional to its depth, $F(z)=-kz$.  Including
gravity, this law predicts the penetration depth for nonzero drop
heights to be $d/d_\circ = (H/d_\circ)^{1/2}$.

To summarize published results for the dependence on drop height,
$d\sim H^{1/3}$ was reported in our first paper \cite{jun},
whereas $d \sim v_\circ \sim h^{1/2}$ was found in
Ref.~\cite{ciamarra} and $d-d_\circ\sim h^{1/2}$ was found in
Ref.~\cite{deBruyn2}. By comparison with the projectile diameter,
the penetrations are shallow in Ref.~\cite{jun} but deep in
Refs.~\cite{ciamarra,deBruyn2}.  Thus there may be no conflict;
the experiments could simply fall into different scaling regimes.
However, evidence of $d\sim H^{1/3}$ for deeper penetrations of
cylinders is reported in \cite{katie}. Furthermore,
Ref.~\cite{deBruyn2} states that our shallow penetration depth
data of \cite{jun} are well-described by their model.  This raises
the possibility that our respective data sets are actually
consistent, and that one of us is mistaken as to the specific
power-law behavior.

The published results for the dependence on projectile density are
also not in agreement.  Our scaling law implies
$d_\circ\sim{\rho_b}^{3/4}$ for $h=0$, Eq.~(\ref{juno}), whereas
$d_\circ \sim \rho_b$ is reported in Ref.~\cite{detlef3}.  Again
the penetrations are more shallow in our work, so the respective
experiments may simply be in different scaling regimes.
Furthermore, our beads are large enough that grain-grain cohesion
is negligible, whereas the grain size and packing fraction are
both considerably smaller in Ref.~\cite{detlef3}. To date, we are
the only group to report density scaling for $h>0$.

Altogether, the results of
Refs.~\cite{jun,ciamarra,deBruyn2,detlef3} suggest that there may
be three distinct sets of impact behavior.  (1) Shallow
penetration into non-cohesive media, where Eq.~(\ref{junlaw})
holds \cite{jun}; (2) Deep penetration into non-cohesive media,
where $d-d_\circ \sim v_\circ$ and $F(v)=-F_\circ-bv$ hold
\cite{ciamarra,deBruyn2}; and (3) penetration into small tenuously
packed grains, where $F(z)=-kz$ holds \cite{detlef3}.

In this paper we (1) provide more details for our original Letter
\cite{jun}, and we (2) report on new experiments designed to
clarify the experimental situation for shallow impacts.  Our
approach is both to improve the reproducibility and accuracy of
the measuring apparatus, and to systematically and widely vary the
drop distance, the ball diameter, and the ball density.  We also
adopt the gas-fluidization preparation method, to see if it
changes the scaling.  We shall demonstrate that the deviations of
data from Eq.~(\ref{junlaw}) are mainly statistical. We shall also
demonstrate that better collapse can be achieved by
Eq.~(\ref{junlaw}) than by the impact speed scaling of
Refs.~\cite{ciamarra,deBruyn2}.  This reaffirms the correctness of
our original work, Refs.~\cite{jun,katie}, and negates the
statement in Ref.~\cite{deBruyn2} that our results can be
described by their model.

\section{Methods}

Our granular medium is P-0140 A-Series technical quality solid
glass spheres from Potters Industries Inc.\ (PA).  The beads are
slightly polydisperse, with a diameter range of 0.25-0.35~mm as
set by US sieve sizes 45-60. The quoted density of the glass
material is 2.5~g/cc.  In our previous work~\cite{jun,katie}, we
poured the beads into a beaker and then gently swirled and tapped
it to achieve a horizontal surface and a random close packing
fraction of about 0.64.  In case this led to irreproducibility of
packing or surface angle, we now prepare the system by
air-fluidization.  The sample container consists of a plexiglass
tube with 8-inch outer diameter, 1/4-inch wall thickness, and 5-ft
height.  The top is open to air, while the bottom consists of a
Brass sieve with 90~$\mu$m mesh opening. Under the sieve is a
windbox consisting of a plexiglass tube of same diameter but
12-inch height.  The glass beads are poured onto the sieve to a
depth of approximately 8 inches.  Dry air is then blown at high
rate into the bottom of the windbox, and up through the glass
beads, until all fines and humidity are removed.  Prior to each
drop, the beads are more gently fluidized, and the airflow is
gradually reduced, so that a flat level surface remains.  It is
crucial to turn down the airflow very slowly, in order to avoid
large gas bubbles that leave behind surface irregularities.  In
earlier work~\cite{ojha} we found that this procedure gives a
packing fraction of $0.590\pm0.004$, as expected for hard
non-cohesive spheres.

We employ two series of spherical projectiles.  The first is
wooden spheres of density $\rho_{b}=0.7$~g/cc and varying
diameter: $D_{b}=\{1/4,\ 1/2,\ 5/8,\ 7/8,\ 1,\ 3/2,\ 2,\
3\}$~inches.  The second is 1-inch diameter spheres of varying
density: hollow polypropylene, 0.28~g/cc; wood, 0.7~g/cc; nylon,
1.2~g/cc; teflon, 2.2~g/cc; ceramic, 3.8~g/cc; steel, 7.9~g/cc;
tungsten carbide (WC), 15~g/cc.  These are held and dropped with
zero speed from the center of the sample tube using a suction
mechanism.  In comparison with our previous work \cite{jun}, the
new sample dimensions and maximum ball diameter are all about
twice as great, but the maximum drop heights are comparable. Since
the penetration depth grows less than linearly with ball diameter,
we judge that sample-size effects are negligible.  Furthermore, we
never observe any grain movement at the edge of the sample as a
result of impact.

The height of the sand surface, the height of bottom of the ball
prior to drop, and the height of the top of the ball after the
drop, are all measured using a micro-telescope (Titan Tool Supply,
Cathetometer TC-11) mounted to a height gauge.  The sample
container and height gauge are approximately 1-foot apart, both
resting on an optical bench.  From the height gauge readings, we
deduce the free-fall height $h$, the penetration depth $d$, and
the total drop distance $H=h+d$.  This method permits study of
penetrations no deeper than the ball diameter, since the top of
the ball must be visible from the side.  For slightly deeper
penetrations, until the ball becomes fully buried, we estimate the
depth from the height of the suction mechanism when it is brought
into contact with the top of ball.

\section{Results}

\begin{figure}
\includegraphics[width=3.00in]{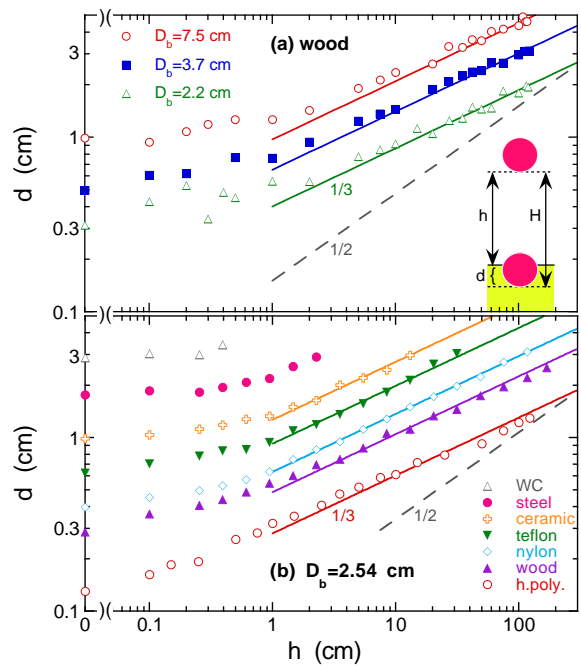}
\caption{Penetration depth $d$ vs free-fall height $h$ for (a)
wooden spheres of different diameter, and (b) one-inch spheres of
different density.  The granular medium is post-fluidized glass
beads of diameter range $0.25-0.35$~mm.  The projectile materials
and densities are as follows: hollow polypropylene, 0.28~g/cc;
wood, 0.7~g/cc; nylon, 1.2~g/cc; teflon, 2.2~g/cc; ceramic,
3.8~g/cc; steel, 7.9~g/cc; tungsten carbide (WC), 15~g/cc.
\label{Raw}}
\end{figure}

Raw data for penetration depth vs free-fall height are displayed
on a log-log plot in Fig.~\ref{Raw} for three example wooden
spheres and for all 1-inch spheres.  The minimum penetration
depth, $d_\circ$ for $h=0$, where the ball bottom was initially
just in contact with the sand surface, is displayed along the left
axis. For decreasing $h$, the penetration depths extrapolate
smoothly to the $h=0$ limit. For increasing $h$, the data appear
to approach a $1/3$ power-law, shown in Fig.~\ref{Raw} by solid
lines.  There is no evidence of $h^{1/2}$ behavior, given by the
dashed lines, which would correspond to the $d\sim v_\circ$
scaling of Ref.~\cite{ciamarra}.  Fig.~\ref{Raw} demonstrates
clearly that $d$ cannot be a power-law over the full range of $h$
due to the nonzero intercept, $d_\circ>0$. One possibility for
simple scaling is that $d$ is a power of total drop distance,
$H=h+d$, as advocated in Ref.~\cite{jun}. Another is that
$d-d_\circ$ is a power of the free-fall distance $h$, as advocated
in Ref.~\cite{deBruyn2}. In the next sections we investigate both
these possibilities.

\section{Total drop distance scaling}

\begin{figure}
\includegraphics[width=3.00in]{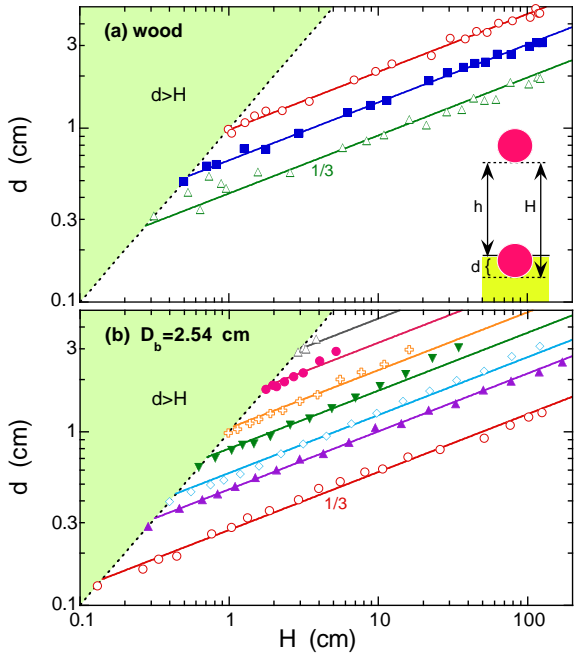}
\caption{Penetration depth $d$ vs total drop distance $H$ for (a) wooden
spheres of different diameter, and (b) one-inch spheres of different
density.  The symbol code is the same as in Figure~1.  The solid
lines are the best fits to $d\propto H^{1/3}$.  Note that the shaded
region, $d>H$, is forbidden.
\label{H}}
\end{figure}

The total drop distance, $H=h+d$, is a relevant parameter because
it relates to the average stopping force via Eq.~(\ref{econ}).
Thus, in Fig.~\ref{H}, we replot all the penetration data of
Fig.~\ref{Raw} vs $H$.  Now the minimum penetration data points,
for free-fall height $h=0$, lie along the line $d=H$; no data may
lie in the shaded region $d>H$.  Whereas in Fig.~\ref{Raw} the
data trended toward $d\sim h^{1/3}$ for large drop heights, now in
Fig.~\ref{H} all the data lie along $d\sim H^{1/3}$ power-laws.
For the lighter spheres, which never submerge, the deviation from
power-law behavior is purely statistical.  For these data sets,
the dynamic range in $H$ is two to three decades, enough to give
confidence in form and a few percent uncertainty in exponent.  For
denser spheres, the $H^{1/3}$ power-law fits gives an acceptable
description, but may deviate for penetrations deeper than about a
ball diameter.

\begin{figure}
\includegraphics[width=3.00in]{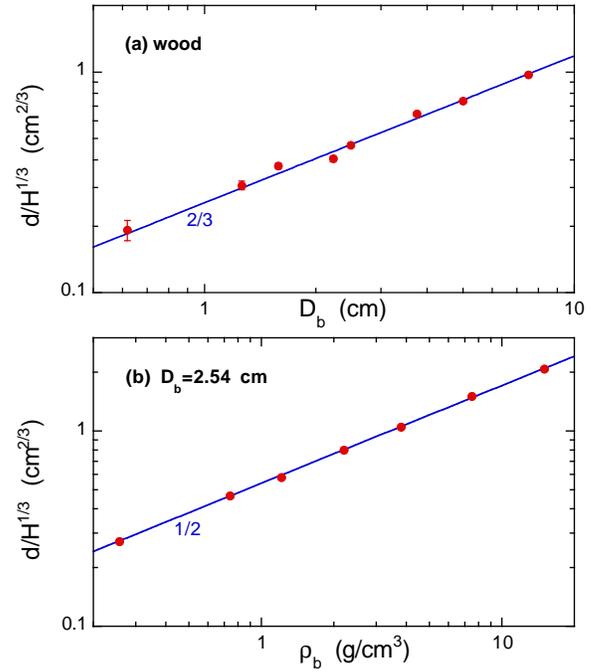}
\caption{Scaled penetration depth vs (a) projectile diameter and
(b) projectile density.  Each point corresponds to a fit to
$d\propto H^{1/3}$ as shown in Fig.~\protect{\ref{H}}.  The solid
lines are the best power-law fits, giving exponents of (a) $2/3$
for diameter and (b) $1/2$ for density.  The inset of (b) shows
the minimum penetration depth $d_\circ$, for $h=0$ and
$D_b=2.54$~cm, as a function of ball density along with power-law
expectations based on Eq.~(\protect{\ref{junlaw}}) (solid blue
line) and on Ref.~\protect{\cite{detlef3}} (dashed green line).
\label{Hscaling}}
\end{figure}

The power-law behavior of Fig.~\ref{H} is further analyzed in
Fig.~\ref{Hscaling}, where we plot the proportionality constant of
the power-law fits, $d/H^{1/3}$, as a function of projectile
properties.  In the top plot, Fig.~\ref{Hscaling}a, we display
results for the wooden spheres vs their diameter.  Note that three
of the points correspond to the three example data sets and fits
shown in Fig.~\ref{H}a.  Evidently, to within statistical
uncertainty, the penetration depth scales as the 2/3 power of
projectile diameter with a dynamic range of slightly over one
decade.  The penetration depth thus scales as
$d\sim{D_{b}}^{2/3}H^{1/3}$. This expression is dimensionally
correct, which suggests that we have empirically uncovered most of
the physics. In other words, the observed ${D_b}^{2/3}$ scaling
lends support to our claim of $H^{1/3}$ scaling.

In Fig.~\ref{Hscaling}b we display the proportionality constant of
the power law fit, $d/H^{1/3}$, for all the one-inch spheres as a
function of their density.  Each point corresponds to one data set
and fit in Fig.~\ref{H}b.  Evidently, to within statistical
uncertainty, the penetration depth scales as the square-root of
projectile density with a dynamic range of over one and one half
decades.  We find the same density dependence as the jet
penetration formula \cite{melosh}.

\begin{figure}
\includegraphics[width=3.00in]{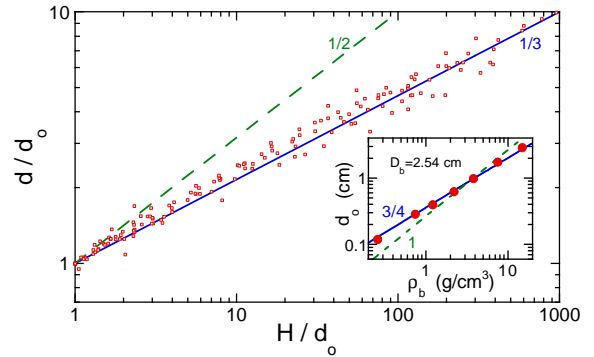}
\caption{All the penetration depth data of
Fig.~\protect{\ref{Raw}}, vs total drop distance, both scaled by
minimum penetration depth. The data all collapse onto
$d/d_\circ=(H/d_\circ)^{1/3}$ (solid blue curve), expected from
Eqs.~(1-2). The characteristic length scale, $d_\circ$, scales
with projectile diameter and the $3/4$ power of projectile density
(inset, solid blue curve). By comparison,
$d/d_\circ=(H/d_\circ)^{1/2}$ and $d_\circ\sim\rho_b$ are
predicted by the model of Ref.~\protect{\cite{detlef3}} (dashed
green curves). \label{DOscaling}}
\end{figure}

Before closing this section, we offer an alternative means of
analyzing penetration data in terms of total drop distance. As
noted in the introduction, Eq.~(\ref{junlaw}) can be recast as
$d/d_\circ = (H/d_\circ)^{1/3}$, where $H=h+d$, $h$ is the
free-fall height, and $d_\circ$ is the minimum penetration depth
for $h=0$.  Thus in Fig.~\ref{DOscaling} we check for data
collapse by plotting $d/d_\circ$ vs $H/d_\circ$, using measured
values of $d_\circ$. The scatter of data is not negligible, but
the average is well-described by the expected $1/3$ power-law,
shown as a solid blue curve.  Even tighter collapse onto this
curve can be achieved if $d_\circ$ is treated as an adjustable
parameter. For comparison, the $d/d_\circ = (H/d_\circ)^{1/2}$
power-law predicted by the model of Ref.~\cite{detlef3} is shown
as a dashed green curve. For both this model and our observations,
$d_\circ$ is the crucial length scale characteristic of a
particular system of projectile and granular medium.  The value of
$d_\circ$ is proportional to the projectile diameter, $D_b$, and a
power of the projectile:medium density ratio, $\rho_b / \rho_g$.
The inset of Fig.~\ref{DOscaling} shows our data for $d_\circ$ vs
$\rho_b$, for all $D_b=2.54$~cm spheres. The results are
consistent with our expectation, $d_\circ\sim\rho_b^{3/4}$
Eq.~(\ref{juno}).  For comparison, the $d_\circ\sim\rho_b$
observation of Ref.~\cite{detlef3} is shown as a solid green
curve. We speculate that the small particle size and the very
tenuous packing in Ref.~\cite{detlef3} are responsible for the
different behavior.

Altogether Figs.~\ref{H}-\ref{DOscaling} show quite convincingly
that the penetration depth scales as
\begin{equation}
    d\sim {\rho_{b}}^{1/2}{D_{b}}^{2/3}H^{1/3},
\label{junlaw2}
\end{equation}
in accord with Eq.~(\ref{junlaw}).  The demonstration here is
stronger than in our prior work, Ref.~\cite{jun}, because the
dynamic ranges are larger and the statistical uncertainties are
smaller.  But more importantly, the demonstration is stronger than
in Ref.~\cite{jun} because here the projectile diameters and
densities are varied more systematically, with one held fixed
while the other is changed. Nonetheless, Ref.~\cite{jun} still
complements the present work in that it established the dependence
of penetration on the properties of the granular medium via
Eq.~(\ref{junlaw}) as $d\sim1/(\mu{\rho_{g}}^{1/2})$.

\section{Impact speed scaling}

\begin{figure}
\includegraphics[width=3.00in]{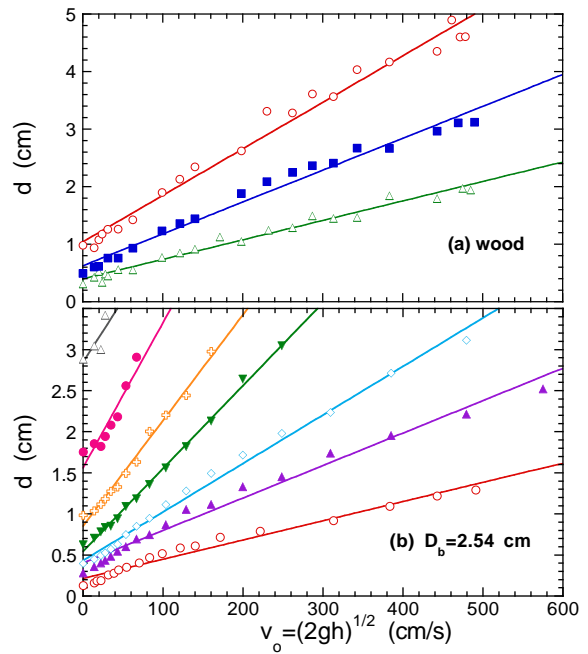}
\caption{Penetration depth vs impact speed for (a) wooden spheres
of different diameter, and (b) one-inch spheres of different
density.  The symbol code is the same as in Figure~1.  The solid
lines are the best fits to $d=d_{\circ v}+v_{\circ}\tau$ where
both $d_{\circ v}$ and $\tau$ are fitting parameters. \label{V}}
\end{figure}

In Refs.~\cite{ciamarra,deBruyn2}, penetration depth data are
reported to scale according to impact speed rather than by
Eq.~(\ref{junlaw}).  Furthermore, Ref.~\cite{deBruyn2} reports
that our earlier data also can be scaled by impact speed.
Therefore in this section we attempt to analyze our new data
similarly.  We begin with Fig.~\ref{V}, where the penetration
depth data of Fig.~\ref{Raw} are replotted as a function of impact
speed, $v_{\circ}=\sqrt{2gh}$.  Contrary to the suggestions of
Refs.~\cite{ciamarra,deBruyn2}, our data do not lie along the best
line fits to $d=d_{\circ v}+v_{\circ}\tau$.  For lighter spheres
the data all curve downwards, while for denser spheres the data
all curve upwards.  Over a subset of speeds, e.g.\
50~cm/s~$<v_{\circ}<$~400~cm/s as in Ref.~\cite{ciamarra}, the fit
to a straight line is satisfactory to within experimental
uncertainty.

\begin{figure}
\includegraphics[width=3.00in]{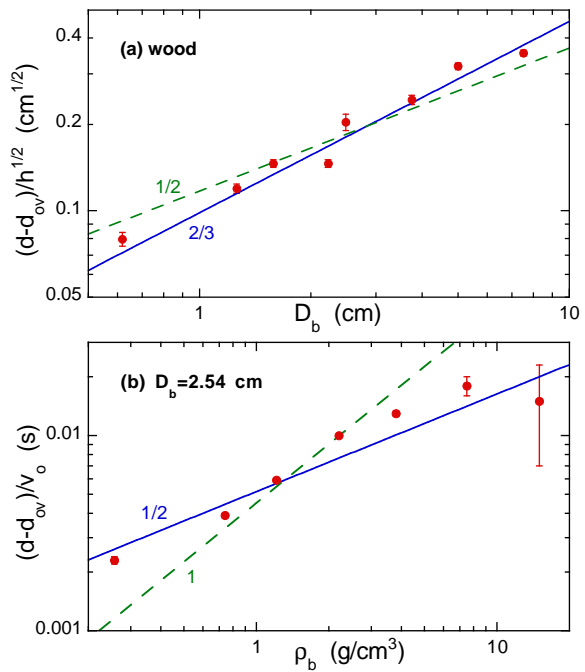}
\caption{Scaled penetration depth vs (a) projectile diameter and
(b) projectile density.  Each point corresponds to a fit to
$d=d_{\circ v}+v_{\circ}\tau$ as shown in Fig.~\protect{\ref{V}}.
The solid lines are the best power-law fits, giving exponents of
(a) $2/3$ for diameter and (b) $1/2$ for density.  The dashed
lines represent dimensionally-simpler expectations, but are not
consistent with the data. \label{Vscaling}}
\end{figure}

To see if we can make sense of the displayed fits to $d=d_{\circ
v}+v_{\circ}\tau$, we examine {\it one} of the fitting parameters
in Fig.~\ref{Vscaling} as a function of projectile properties. The
top plot in shows $(d-d_{\circ v})/h^{1/2}$, i.e.\ the fitting
parameter $\tau$ times $\sqrt{2g}$, as a function of ball
diameter.  A reasonable power-law fit can be made to
${D_{b}}^{2/3}$, which curiously is the same diameter exponent as
Eq.~(\ref{junlaw}). Though this fit gives a fine description, it
is not dimensionally simple.  If $(d-d_{\circ v})\propto
{D_{b}}^{2/3}h^{1/2}$ is true, then there must be another
important length scale in the problem that enters the
proportionality constant.  It would have been simpler had we found
$(d-d_{\circ v})/h^{1/2}\propto {D_{b}}^{1/2}$, shown by a dashed
line. Such behavior clearly differs from the data.

The bottom plot of Fig.~\ref{Vscaling} shows the fitting parameter
$\tau=(d-d_{\circ v})/v_{\circ}$ as a function of ball density.
The results curve downwards, and cannot be very well described by
a power-law.  Nonetheless, the best power-law fit would be to
${\rho_{b}}^{1/2}$, as shown by the solid line.  Curiously, the
density exponent is the same as in Eq.~(\ref{junlaw}).  If true,
this corresponds to scaling with the square root of impact energy,
$(d-d_{\circ v})\sim\sqrt{\rho_{b}{v_{\circ}}^{2}}$.  Scaling with
impact momentum, suggested in the abstract of Ref.\cite{deBruyn2},
would correspond to $(d-d_{\circ v})\sim{\rho_{b}}v_{\circ}$ as
shown by the dashed line. Such behavior is vastly different from
the data.  Contrary to the abstract, however, the final scaling
advocated in Ref.~\cite{deBruyn2} is $(d-d_{\circ
v})\sim{\rho_{b}}^{1/2}v_{\circ}$.  This corresponds to the solid
curve in Fig.~\ref{Vscaling}b, which still is not a satisfactory
fit.  Even better powerlaw fits can be made in both
Figs.~\ref{Vscaling}a-b if the last point is omitted; however, the
resulting exponents do not lead to dimensionally simple scaling.
While Figs.~\ref{V}-\ref{Vscaling} alone do not unequivocally rule
out scaling by impact speed, the contrast with scaling by total
drop distance in Figs.~\ref{H}-\ref{Hscaling} is striking.

Altogether, the best description of our new data in terms of impact
speed is
\begin{equation}
    d-d_{\circ v}\propto {\rho_{b}}^{1/2}{D_{b}}^{2/3}h^{1/2},
\label{johnlaw}
\end{equation}
where the intercept, $d_{\circ v}>0$, is a free fitting parameter
as yet unaccounted for.  According to the model of
Ref.~\cite{deBruyn2}, the intercept can be explained by a yield
stress but only if it is negative, which is not the case for our
experiments.  If Eq.~(\ref{johnlaw}) is true, the combined density
and free-fall height dependencies imply that impact energy is the
crucial parameter, not the impact momentum. For a complete
understanding, one would still have to account for both the free
parameter $d_{\circ v}$ as well as an additional length scale in
the proportionality constant.

\section{Data Collapse}

\begin{figure}
\includegraphics[width=3.00in]{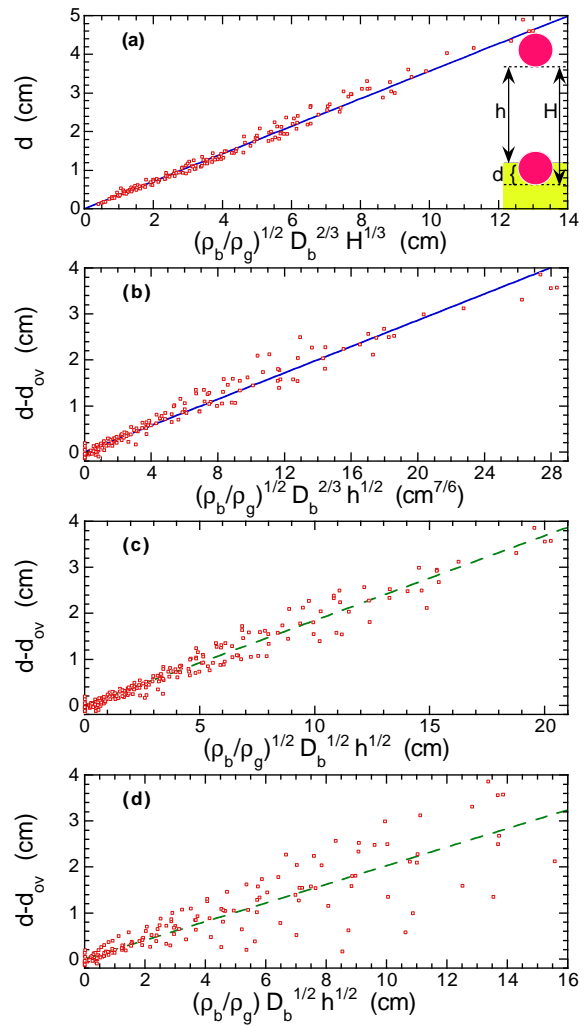}
\caption{Penetration depth vs (a) scaled total drop distance and
(b-d) scaled free-fall height.  In (a), the exponents for sphere
density, sphere diameter, and total drop distance are taken from
the fits of Figs.~\protect{\ref{H}}-\protect{\ref{Hscaling}}.  In
(b), the exponents are taken from the fits of
Figs.~\protect{\ref{V}}-\protect{\ref{Vscaling}}.  In (c), the
exponents are taken according to the final prescription of
Ref.~\protect{\cite{deBruyn2}}.  In (d), the exponents are taken
such that the x-axis is a length scale proportional to impact
momentum. In (b-d), $d_{\circ v}$ is a fitting parameter from
Fig.~\protect{\ref{V}} that is systematically different from the
penetration depth at zero impact speed. \label{Collapse}}
\end{figure}

As an alternative means to compare the relative quality of
candidate scaling descriptions, we now attempt to collapse the
penetration depth data of Fig.~\ref{Raw} according to both the
total drop distance $H$ as well as according to the impact speed
$v_{\circ}=\sqrt{2gh}$.  For collapse via Eq.~(\ref{junlaw2}), we
plot penetration depths vs
$(\rho_{b}/\rho_{g})^{1/2}{D_{b}}^{2/3}H^{1/3}$ in
Fig.~\ref{Collapse}a.  Though the grain density has not been
explicitly varied here, we assume the same dependence as observed
in Ref.~\cite{jun}; this renders the $x$-axis dimensionally
correct. Evidently, in Fig.~\ref{Collapse}a, the data collapse
tightly to a straight line with a statistical deviation that is
roughly a constant percentage. Assuming a proportionality constant
of $0.14/\mu$, as in Eq.~(\ref{junlaw}), we find that the repose
angle of the post-fluidized glass beads is
$\theta_{r}=21^{\circ}$.  This is slightly smaller than the value
measured in Ref.~\cite{jun} for glass beads at random close
packing, $\theta_{r}=24^{\circ}$, as expected.

For collapse via Eq.~(\ref{johnlaw}), we subtract the {\it fitted}
intercept from the penetration depths and plot vs
$(\rho_{b}/\rho_{g})^{1/2}{D_{b}}^{2/3}h^{1/2}$ in
Fig.~\ref{Collapse}b.  The degree of collapse is noticeably not as
tight as in Fig.~\ref{Collapse}a, with a percentage deviation that
blooms for smaller penetrations. Relatedly, the systematic
curvature of the data away from $(d-d_{\circ v})\propto h^{1/2}$,
seen in Fig.\ref{V}b, is reflected here by a deviation of the
actual penetration depth for $h=0$ from the value of the fitting
parameter $d_{\circ v}$.  If instead we plot $d-d_{\circ}$, where
$d_{\circ}$ is the actual observed penetration for $h=0$, then the
degree of collapse in Fig.~\ref{Collapse}b is notably worse. Also,
consistent with the diameter and density dependence shown in
Fig.~\ref{Vscaling}, the degree of collapse is worse in
Fig.~\ref{Collapse}c when the penetration depth is plotted vs
$(\rho_{b}/\rho_{g})^{1/2}{D_{b}}^{1/2}h^{1/2}$, the final scaling
advocated in Ref.~\cite{deBruyn2}.

Finally, in Fig.~\ref{Collapse}d, we make one last attempt at
collapsing our data.  The abstract of Ref.~\cite{deBruyn2} states
that ``...the penetration depth of the spheres increases linearly
with the incident momentum of the projectile, but with a
zero-momentum intercept that can be positive or negative."
According to this prescription, we subtract the fitted intercept
and plot the otherwise-raw penetration depth data of
Fig.~\ref{Raw} vs the dimensionally-simplest quantity proportional
to momentum: $(\rho_{b}/\rho_{g}){D_{b}}^{1/2}h^{1/2}$.  This
gives a nearly-random scattering of data points without the least
hint collapse.  Thus, impact momentum does not determine the
penetration depth for our data.

\section{Conclusion}

Our new data for the shallow penetration of spheres into a loose
granular medium strongly support our previous conclusions,
Eq.~(\ref{junlaw}), for the scaling of penetration depth. By
improving preparation reproducibility and measurement accuracy, we
demonstrate that the depth scales as the $1/3$ power of the total
drop distance $H$. In particular there is a nonzero penetration
even for zero drop height $h$, where the penetration depth equals
the total drop distance.  By systematically varying the projectile
diameter at fixed density, and by systematically varying the
projectile density at fixed diameter, we demonstrate that the
depth scales as the $1/2$ power of projectile density and the
$2/3$ power of projectile diameter.  And by changing the sample
preparation from random close packing in Ref.~\cite{jun} to a
random loose packing here, we demonstrate that sample preparation
plays no crucial role.  As long as the medium is loose and
noncohesive, Eq.~(\ref{junlaw}) should apply though with a value
of $\mu$ that reflects the packing state.  The burning question is
now the nature of the granular mechanics that gives rise to this
reaffirmed scaling behavior.  The force law cannot be as suggested
in Ref.~\cite{deBruyn2}, where the impact momentum of the
projectile and a yield stress for the granular medium are crucial
inputs. The positive intercept for penetration depth vs drop
height, and the scaling with total drop distance $H$ rather than
with free-fall height $h$, suggest instead that the stopping force
may depend on the projectile's instantaneous depth as well as its
speed.  Such a scenario would be more in line with recent reports
of the hydrostatic-like nature of the force on an object moving
horizontally \cite{schiffer01} or vertically
\cite{schiffer04,detlef3} through a granular medium.

{\it Acknowledgements} This material is based upon work supported
by the National Science Foundation under grant DMR-0305106.  We
thank J.R.\ de~Bruyn and D.\ Lohse for helpful conversations.

\bibliography{CraterRefs}

\end{document}